\documentclass[11pt]{article} 
\usepackage{amsfonts,amscd,amsmath,graphicx}

\def\NP#1#2{ Nucl.Phys. B#1 (#2)} 
\def\PL#1#2{ Phys.Lett. B#1 (#2)}
\def\CMP#1#2{ Commun.Math.Phys. #1 (#2)} 
\def\MPL#1#2{ Mod.Phys.Lett.A#1 (#2)} 
 
\def\PR#1#2{Phys.Rev. D#1 (#2)} 
\def\IJMP#1#2{ Int.J.Mod.Phys. A#1 (#2)}
\def\ATMP#1#2{ Adv.Theor.Math.Phys. #1 (#2)}

\newcommand{\So}{SL(2,R)} 
\newcommand{\Os}{OSp(1\vert 2)}
\newcommand{\Sl}{SL(2\vert 1)} 
\newcommand{\so}{\hat{sl}(2)}
\newcommand{\os}{\hat{osp}(1\vert 2)}
\newcommand{\ssl}{\hat{sl}(2\vert 1)} 
\newcommand{\su}{\hat{su}(2)}
\newcommand{\zo}{z_{12}} 
\newcommand{\zt}{z_{12}^2}
\newcommand{\pd}{\partial} 
\newcommand{\oh}{\frac{1}{2}}
\newcommand{\oq}{\frac{1}{4}} 
\newcommand{\dpm}{\dot\pm}
\newcommand{\pp}{\dot +} 
\newcommand{\mm}{\dot -}
\newcommand{\te}{\tilde\eta} 
\newcommand{\txi}{\tilde\xi}
\newcommand{\bz}{\bar z} 
\newcommand{\sif}{\sigma^1}
\newcommand{\sis}{\sigma^2} 
\newcommand{\ep}{\text e}
\newcommand{\zp}{z^{\prime}} 
\newcommand{\tp}{\tilde p}
\newcommand{\tga}{\tilde\gamma}

\title{On Affine Lie Superalgebras, AdS{}$_3$/CFT Correspondence And
  World-Sheets For World-Sheets}
\author{Oleg Andreev\thanks{E-mail: andre@landau.ac.ru}\\ \\
  L.D.Landau Institute for Theoretical Physics,\\
  Kosygina 2, 117334 Moscow, Russia} \date{}
\begin{document}
\maketitle

\vspace{-8cm}
\begin{flushright}hep-th/9901118 \\
  LANDAU-99/HEP-A1
\end{flushright}

\vspace{6cm}
\begin{abstract}
  Space-time $N=2$ and $N=4$ superconformal algebras can be built
  using world-sheet free fields or, equivalently, free field
  representations of $\os$ and $\ssl$, respectively. A prescription
  for the calculation of space-time correlators is given. As
  applications we compute one and two-point correlation functions of
  Virasoro (superVirasoro) generators. We also present a possible
  scenario for taking into account of NS fivebranes in the framework
  of our construction for the description of superstring on $AdS_3$.
  It is nothing but a simplified version of Green's ``world-sheets for
  world-sheets''.
\end{abstract}


\section{Introduction}
\renewcommand{\theequation}{1.\arabic{equation}}
\setcounter{equation}{0} 
Several promising ideas are motivations for this paper:

(i) The newest, and perhaps most interesting, of the dualities of
string theory is the one conjectured by Maldacena, which relates the
large N expansion of gauge theories (conformal field theories) in $d$
dimensions to string theory on the product of $d+1$-dimensional
Anti-de Sitter ($AdS$) space with a compact manifold \cite{Mal}.

(i') In fact, the AdS/CFT correspondence is an example of the 't Hooft
Holographic principle according to which a quantum theory with gravity
must be described by a boundary theory \cite{H,LS}.

(ii) The Green's idea is that the two-dimensional world-sheet underlying
the dynamics of string theory may itself emerge as the embedding space
of a two-dimensional string theory which has a more primitive
structure \cite{Green}.

(iii) In general, a supersymmetry algebra contains central charges
that couple to topological quantum numbers of solitons \cite{SUSY}. In
the stringy context these charges are carried by p-branes.

Each of these ideas has its own attractive features, so it serves as a
subject for numerous investigations. In particular, the possible
relation of string theory on $AdS_3$ to conformal field theory has
been a subject of much interest; see [6, 8-13] for discussions
motivated by recent developments, and additional references. It is
amusing, however, that all these ideas can be found as elements of a
construction which is superstring theory on $AdS_3$. It is clear that
applying the ideas together should help us with understanding such
construction, so it will be the purpose of this paper.

World-sheet description of superstring theory on $AdS_3$ has been
initiated by Giveon, Kutasov and Seiberg \cite{GKS} who started with
the bosonic part of the world-sheet Lagrangian or, what is equivalent,
with the $G_{\mu\nu}\pd X^\mu\pd X^\nu$ term of the corresponding
$\sigma$ model\footnote{In fact, this is an old story. Such action was
  also considered by Gawedzki about ten years ago within his path
  integral approach to the SL(2) Wess-Zumino-Witten model \cite{Gaw}.}
\begin{equation}
{\cal L}\sim \pd\varphi\bar\pd\varphi+\ep^{2\varphi}
\pd\bar\gamma\bar\pd\gamma
\end{equation}
and transformed it into a form corresponding to the Wakimoto free
field representation of the SL(2) Wess-Zumino-Witten model. In this
case the interaction term becomes the screening operator of the $\so$
algebra.  The $N=1$ superconformal algebra on the world-sheet is
achieved by introducing free fermions that are lower components of
superfields.  They also succeeded in finding the global $N=4$
superconformal algebra for the Neveu-Schwarz (NS) sector in
space-time.

In spite of certain progress in understanding string propagation on
$AdS_3$ the description of Giveon and co-workers is subjected to some
drawbacks. One of them is that space-time $N=4$ algebra (boundary
algebra) is not realized in a manifestly supersymmetric way. Moreover,
the use of the standard procedure for building fermionic vertices by
\cite{FMS} does not allow one to explicitly write down all the
fermionic generators of the algebra. Recently Ito proposed a simple
remedy\footnote{Due to its manifest space-time supersymmetry it looks
  like an analog of the Green-Schwarz formulation of superstring for
  the problem at hand, so it seems reasonable to pay more attention to
  this analogy in future research.} to improve the situation. He based
on the Wakimoto free field representations of some affine Lie
superalgebras \cite{I}. His construction provides the $N=1,\,2,\,4$
superconformal algebras in space-time through the world-sheet 
$\os,\,\ssl ,\,\hat{sl}(2\vert 2)$ affine superalgebras, respectively
\footnote{Note that the relations between affine and superconformal
  algebras $\os\rightarrow N=1,\,\,\ssl\rightarrow N=2,\,\,
  \hat{sl}(2\vert 2)\rightarrow N=4$ are also known from the
  Drinfeld-Sokolov reduction, which is an entirely different topic.}.

However, the construction discussed in \cite{GKS,I} remains
inconsistent because, while the Holographic principle says that string
theory on $AdS_3$ is described by a boundary theory, the boundary
$N=4$ algebra is built in terms of all $AdS_3$ coordinates rather than
the boundary ones only; improving this will be the first goal of the
present paper. Moreover, the realization of the number of NS5-branes
as the level of the world-sheet $\so$ algebra does not look attractive
too. The point is that it does not allow one to interpret the central
charge of the boundary algebra in a purely topological way (from the
world-sheet point of view). So, this will be our second goal. At the
present time it is not known whether these goals may be achieved. So
we are bound to learn something if we succeed.

The outline of the paper is as follows. In section 2 we describe a
formal construction for lifting bosonized affine Lie algebras on the
world-sheet into two-dimensional space-time. We will not only
reproduce the known results, but find rather amusing relations between
$N=2,\,4$ superconformal algebras and affine Lie superalgebras $\os$,
$\ssl$. In particular, such 2d space-time may be considered as the
boundary of $AdS_3$. Section 3 will present some applications to
string on $AdS_3$. We give a prescription for the calculation of
space-time correlators and possible mechanism for taking into account
NS5-branes in a purely topological way. Section 4 will present the
conclusions and directions for future work. In the appendices we give
technical details which are relevant for the explicit construction of
the affine Lie superalgebras in space-time and computation of some
correlators.


\section{Superconformal Algebras in Space-Time}
\renewcommand{\theequation}{2.\arabic{equation}}
\setcounter{equation}{0} We now turn to the problem of the
correspondence between symmetries on the world-sheet and
two-dimensional space-time. In particular, the latter may be the
boundary of $AdS_3$. In this case one can consider a construction by
starting with the Euclidean version of $AdS_3$ whose coordinates are
$(\varphi,\gamma,\bar\gamma)$ and restricting the analysis to the
problem of building superconformal algebras in space-time which is the
boundary of $AdS_3$ with coordinates $(\gamma,\bar\gamma)$ (see
subsection 3.1.4 for details). Thanks to the underlying world-sheet
structure such construction may be called the stringy representation
of affine superalgebras. However, before we give examples, it might be
useful to present some mathematical background.

Let $P_0$ be a point on the world-sheet which is a Riemann surface and
$(U_0,z)$ be a coordinate patch such that $P_0\in U_0$ and $z=0$ for
$P_0$. Let $P_\gamma$ be a point on another two-dimensional manifold
(space-time) and $(U_\gamma,\gamma)$ be a proper coordinate patch.
Let there is a map $\gamma :\,\, U_0\rightarrow U_\gamma$, i.e. the
local embedding of the world-sheet into space-time. It is useful to
classify such maps by their degrees, i.e.  integers $p$ which are
given by
\begin{equation}\label{top}
  p=\oint_{C_0} dz\,\pd\gamma\gamma^{-1}\,(z)\quad,
\end{equation}
where the contour $C_0$ surrounds $0$. Thanks to the local
construction $C_0$ is contractible on the world-sheet. The
normalization is fixed by setting $p=1$ for $\gamma (z)=z$ i.e.,
$\oint_{C_0}\frac{dz}{z}=1$. In the quantum case $p$ is interpreted as
the topological charge\footnote{In the stringy context $p$ is
  interpreted as the number of infinitely stretched fundamental
  strings at the boundary of $AdS_3$ \cite{GKS}. We will see in
  section 3 that there is also another interpretation.}.

\subsection{The Virasoro case}
As a warmup, let us consider the Virasoro algebra in space-time and,
as by-product, reproduce the result of \cite{GKS}. The most obvious
way to do this is the following.  Consider a differential operator
realization of the Virasoro algebra. By substitutions
\begin{equation*}
\pd_\gamma\rightarrow\beta(z)\quad ,
\quad\gamma\rightarrow\gamma(z)\quad,
\end{equation*}
where $\beta(z),\,\gamma(z)$ are now the quantum fields, we get the
quantum algebra. This is the standard machinery for building the
Wakimoto free field representation within 2d conformal field theories.

It is well known that the differential operator realization for the
centerless Virasoro algebra is given by
\begin{equation}\label{cV}
  L_n=-\gamma^{n+1}\pd_\gamma\quad,\quad n\in\text{\bf Z} \quad.
\end{equation}

We then take the Virasoro generators to be (normal ordering is
implied here and below)
\begin{equation}\label{qV}
  L_n=\oint_{C_0} dz\,-\gamma^{n+1}\beta(z)\quad.
\end{equation}

Using the standard techniques (see, e.g., \cite{FMS}) together with
the Operator Product (OP) expansion
$\beta(z_1)\gamma(z_2)=\frac{1}{\zo}+O(1)$ we
find\footnote{$[L_n,c]$ can be shown to be zero.}
\begin{equation}\label{Vir}
  [L_n,L_m]=(n-m)L_{n+m}+\frac{c}{12}(n^3-n)\delta_{n+m,0}\quad,
\end{equation}
where the cental charge is $c=12p$.

Now, let us assume that there is one more free field on the
world-sheet. Let it be a scalar field $\varphi$ whose OP expansion is
$\varphi(z_1)\varphi(z_2)=-\ln\zo +O(1)$. The most obvious
generalization of what we have just done is
\begin{equation}\label{qV2}
  L_n=\oint_{C_0} dz\,-\gamma^{n+1}\beta+f(n)\gamma^n\pd\varphi(z)\quad.
\end{equation}
A simple algebra shows that such defined $L_n$'s obey \eqref{Vir} only
for $f(n)=(n+1)f_0$ with $f_0=\text{const}$. This improved
construction gives the central charge $c=12(1+f_0^2)p$.

On the other hand, the set of the free fields we used is exactly what
one needs to build the Wakimoto free field representation of the $\so$
algebra.  We refer to Appendix A for more details. With $\beta$ and
$\pd\varphi$ as in \eqref{fso} and normal ordering implied in
\eqref{qV2}, the formula for $L_n$ is modified to
\begin{equation}\label{Vso}
  L_n=\oint_{C_0} dz\,-(1+2\frac{f_0}{\alpha_+}(n+1))\gamma^{n+1}J^-+
  2\frac{f_0}{\alpha_+}(n+1)\gamma^nJ^0(z)\quad.
\end{equation} 
It is easy to see that the result found in \cite{GKS} is recovered by
inserting $f_0=-\oh\alpha_+$ into this formula. As a result, the
central charge is $c=6kp$. We will return to a discussion of this
point in section 3.

\subsection{The $N=2$ case}
The generalization of the above construction is obtained simply by
adding fermionic coordinates, which makes the space-time supersymmetry
manifest. In the simplest case that we will consider we add a
fermionic coordinate $\xi$ that geometrically turns space-time into a
superspace. In general one would expect the $N=1$ superconformal
structure for this case. Happily, this is not the whole story. It is
easy to see that the following differential operators
\begin{equation}\label{n2}
  L_n=-\gamma^{n+1}\pd_\gamma-\frac{n+1}{2}\gamma^n\xi\pd_\xi\quad,\quad
  T_n=\gamma^n\xi\pd_\xi\quad,\quad 
  G_r^+=\gamma^{r+\oh}\xi\pd_\gamma\quad,\quad
  G_r^-=-\gamma^{r+\oh}\pd_\xi\quad,
\end{equation}
with $n,\,r\in\text{\bf Z}\,(\text{\bf Z}+\oh)$, provide a
differential operator realization of the centerless $N=2$
superconformal algebra.

By substitutions
\begin{equation*}
\pd_\gamma\rightarrow\beta(z)\quad,\quad\gamma\rightarrow\gamma(z)\quad,\quad
  \pd_\xi\rightarrow\eta(z)\quad,\quad\xi\rightarrow\xi(z)\quad, 
\end{equation*}
where $(\beta(z),\gamma(z))$ and $(\eta(z),\xi(z))$ are now the
quantum fields, we can obtain the leading terms in the quantum
generators.

Finally, we can compute the ``exact'' generators, getting for
instance
\begin{equation}\label{qn2}
  \begin{split}
    L_n&=\oint_{C_0}dz\,-\gamma^{n+1}\beta-\frac{n+1}{2}\gamma^n\xi\eta+
    i\frac{n+1}{2}\pd\varphi\gamma^n(z)\quad,\\
    T_n&=\oint_{C_0}dz\,\gamma^n\xi\eta+i\pd\varphi\gamma^n(z)\quad,\\
    G^+_r&=\oint_{C_0}dz\,\gamma^{r+\oh}\xi\beta
    -i(r+\oh)\pd\varphi\xi\gamma^{r-\oh}(z)\quad,\\
    G_r^-&=\oint_{C_0}dz\,-\gamma^{r+\oh}\eta(z)\quad.
  \end{split}
\end{equation}
To define the generators in the Ramond and topological sectors we
need rational powers of $\gamma$. This can be done either by analytic
continuation or by bosonization. We will return to this point later in
section 3.

Conformal techniques may of course also be used to calculate the
commutation relations of these generators, although the algebra is
more involved. By using the OP expansion
$\eta(z_1)\xi(z_2)=\frac{1}{\zo}+O(1)$, we find
\begin{equation}\label{N2}
\begin{split}
    [L_n,L_m]&=(n-m)L_{n+m}+\frac{c}{12}(n^3-n)\delta_{n+m,0}\quad,\quad
    [L_n,T_m]=-mT_{n+m}\quad,\\
    [T_n,T_m]&=\frac{c}{3}n\delta_{n+m,0}\quad,\quad
    [L_n,G_r^\pm]=(\frac{n}{2}-r)G^\pm_{r+n}\quad,\quad
    [T_n,G^\pm_r]=\pm G^\pm_{r+n}\quad,\\
    \{G^+_r,G_s^-\}&=L_{r+s}+\oh(r-s)T_{r+s}+\frac{c}{6}(r^2-\oq
    )\delta_{r+s,0} \quad,\quad\{G^\pm_r,G_s^\pm\}=0\quad,
  \end{split}
\end{equation}
where $c=6p$.

This time the set of the free fields we used is exactly what one needs
to build the Wakimoto free field representation of the $\os$ algebra.
We refer to Appendix A for more details. With $\beta$, $\eta$ and
$\pd\varphi$ as in \eqref{fos} and normal ordering implied in
\eqref{qn2}, the generators are thus
\begin{equation}\label{qn2os}
  \begin{split}
    L_n&=\oint_{C_0}dz\,-(1+i\,\frac{n+1}{2\alpha_+})\gamma^{n+1}J^-
    +i\,\frac{n+1}{2\alpha_+}\gamma^nJ^0
    -\frac{n+1}{2}(1+\frac{i}{2\alpha_+})\gamma^n\xi j^-(z)\quad,\\
    T_n&=\oint_{C_0}dz\,-\frac{i}{\alpha_+}\gamma^{n+1}J^-
    +\frac{i}{\alpha_+}\gamma^nJ^0+(1-\frac{i}{2\alpha_+})\xi\gamma^nj^-(z)
\quad,\\
    G^+_r&=\oint_{C_0}dz\,(1+i(r+\oh))\xi\gamma^{r+\oh}J^-
    -i(r+\oh)\xi\gamma^{r-\oh}J^0(z)\quad,\\
    G_r^-&=\oint_{C_0}dz\,-\gamma^{r+\oh}j^--\oq\xi\gamma^{r+\oh}J^-(z)\quad.
  \end{split}
\end{equation}

At this point, it is necessary to make a remark. It is also not
difficult to build the $N=1$ superconformal algebra starting from the
following differential operators (subset of \eqref{n2})
\begin{equation}\label{n1-2}
  L_n=-\gamma^{n+1}\pd_\gamma-\frac{n+1}{2}\gamma^n\xi\pd_\xi   
  \quad,\quad
  G_r=a\gamma^{r+\oh}\xi\pd_\gamma-b\gamma^{r+\oh}\pd_\xi
  \quad,\quad
  \text{with}\quad ab=1\quad.
\end{equation}
The analysis proceeds as in above and leads to the $N=1$
superconformal algebra with an arbitrary central charge. For a special
case we recover the result by Ito \cite{I}.

\subsection{The $N=4$ case}
Having the $N=2$ superconformal algebra, we now wish to realize the
$N=4$ superconformal algebra in a similar way. First, by adding two
fermionic coordinates $\xi,\txi$, we turn space-time into a $N=2$
superspace. The first order fermionic differential operators which can
be constructed out of $(\gamma,\,\xi,\,\txi )$ are
\begin{equation*}
  \gamma^a\pd_{\xi}
  \quad,\quad
  \gamma^{\tilde a}\pd_{\txi}
  \quad,\quad
  \gamma^b\xi\pd_\gamma
  \quad,\quad
  \gamma^{\tilde b}\txi\pd_\gamma
  \quad,\quad
  \xi\txi\gamma^c\pd_{\txi}
  \quad,\quad
  \txi\xi\gamma^{\tilde c}\pd_{\xi}
  \quad,
\end{equation*}
i.e., we have two more operators than it is necessary for $N=4$. Since
we are interested in $N=4$, we omit a general analysis of what we can
build from this set of operators. As to the differential operator
realization of the centerless $N=4$ superconformal algebra, it is
given by
\begin{equation}\label{n4}
  \begin{split}
    L_n&=-\gamma^{n+1}\pd_\gamma
    -\frac{n+1}{2}\gamma^n(\xi\pd_\xi+\txi\pd_{\txi} )
    \quad,\quad \\
    T_n^0&=\oh\gamma^n(\xi\pd_\xi-\txi\pd_{\txi}) \quad,\quad
    T_n^-=\gamma^n\txi\pd_\xi \quad,\quad T_n^+=\gamma^n\xi\pd_{\txi}
    \quad,\\
    G^+_r&=i\sqrt2\gamma^{r+\oh }\pd_{\txi} \quad,\quad\quad\quad
    G^{\pp}_r=i\sqrt2 \Bigl(\gamma^{r+\oh }\xi\pd_\gamma +(r+\oh
    )\xi\txi\gamma^{r-\oh }\pd_{\txi}\Bigr)
    \quad,\quad \\
    G^{\mm}_r&=i\sqrt2\gamma^{r+\oh }\pd_\xi \quad,\quad\quad\quad
    G^-_r=i\sqrt2 \Bigl(\gamma^{r+\oh } \txi\pd_\gamma +(r+\oh
    )\txi\xi\gamma^{r-\oh }\pd_{\xi}\Bigr) \quad,
  \end{split}
\end{equation}
where $n,\,r\in\text{\bf Z}\,(\text{\bf Z}+\oh )$.

By substitutions
\begin{equation*}
  \pd_\gamma\rightarrow\beta(z)\quad,\quad
  \gamma\rightarrow\gamma(z)\quad,\quad
  \pd_\xi\rightarrow\eta(z)\quad,\quad
  \xi\rightarrow\xi(z)\quad,\quad 
  \pd_{\txi}\rightarrow\te (z)\quad,\quad
  \txi\rightarrow\txi (z)\quad,
\end{equation*}
where $(\beta(z),\gamma(z))$, $(\eta(z),\xi(z))$ and $(\te (z),\txi
(z))$ are now the quantum fields, we can assume leading terms in
quantum generators. It is amusing that they turn out to be
exact!$\,$\footnote{We will discuss this point later in section 3.}
Indeed, a simple algebra shows that the following generators
\begin{equation}\label{qn4}
  \begin{split}
    L_n&=\oint_{C_0}dz\,-\gamma^{n+1}\beta
    -\frac{n+1}{2}\gamma^n(\xi\eta+\txi\te )\,(z)
    \quad,\quad \\
    T_n^0&=\oh\oint_{C_0}dz\,\gamma^n(\xi\eta-\txi\te )\,(z)
    \quad,\quad T_n^-=\oint_{C_0}dz\,\gamma^n\txi\eta\,(z) \quad,\quad
    T_n^+=\oint_{C_0}dz\,\gamma^n\xi\te\,(z)
    \quad,\\
    G^+_r&=i\sqrt2\oint_{C_0}dz\,\gamma^{r+\oh }\te\,(z)
    \quad,\quad\quad\,\,\, G^{\pp}_r=i\sqrt2\oint_{C_0}dz\,
    \gamma^{r+\oh }\xi\beta +(r+\oh )\xi\txi\gamma^{r-\oh }\te\,(z)
    \quad,\quad \\
    G^{\mm}_r&=i\sqrt2\oint_{C_0}dz\,\gamma^{r+\oh }\eta\,(z)
    \quad,\quad\quad\,\,\, G^-_r=i\sqrt2 \oint_{C_0}dz\,\gamma^{r+\oh
      } \txi\beta+(r+\oh )\txi\xi\gamma^{r-\oh }\eta\,(z)
  \end{split}
\end{equation}
satisfy the $N=4$ superconformal algebra with $c=6p$, namely
\begin{equation}\label{N4}
  \begin{split}
    [L_n,L_m]&=(n-m)L_{n+m}+\frac{c}{12}(n^3-n)\delta_{n+m,0}
    \quad,\\
    [T_n^a,T_m^b]&=\frac{c}{12}ng^{ab}\delta_{n+m,0}+f^{ab}_cT^c_{n+m}
    \quad,\\
    \{G^\alpha_r,G_s^\beta\}&= \frac{c}{3}(r^2-\oq
    )\delta^{\alpha\beta}\delta_{r+s,0}+ 2\delta^{\alpha\beta}L_{r+s}+
    4(r-s)(\sigma^a)^{\alpha\beta}T_{r+s}^bg_{ab}
    \quad,\\
    [L_n,T_m^a]&=-mT_{n+m}^a \quad,\quad
    [L_n,G_r^\alpha]=(\frac{n}{2}-r)G^\alpha_{r+n} \quad,\quad
    [T_n^a,G^\alpha_r]=(\sigma^a)^\alpha_\beta G^\beta_{r+n} \quad.
  \end{split}
\end{equation}
Here $g^{00}=1,\,\,g^{+-}=g^{-+}=2;\,\,f^{0+}_+=
f^{-0}_-=1,\,\,f^{+-}_0=2;\,\,a,b,c=0,+,-$; $\delta^{+-}=\delta^{\pp
  ,\mm }=1$, $\alpha,\,\beta=+,-,\pp ,\mm $;
$(\sigma^a)^{\alpha\beta}=(\sigma^a)^{\alpha}_\gamma
\varepsilon^{\gamma\beta}$, where $\varepsilon$ is the antisymmetric
tensor with $\varepsilon^{+-}=\varepsilon^{\mm \pp}=1$; and the
tensors $(\sigma^a)^\alpha_\beta$ are in the representation
\begin{equation*}
  (\sigma^+)^-_{\pp }=(\sigma^-)^{\pp }_-=-(\sigma^+)^{\mm }_+=
  -(\sigma^-)^+_{\mm }=1\quad,\quad
  (\sigma^0)^+_+=(\sigma^0)^{\pp }_{\pp }=
  -(\sigma^0)^-_-=-(\sigma^0)^{\mm }_{\mm }=\oh\quad.
\end{equation*}

In contrast to the previous examples, the set of the free fields used
to build the $N=4$ superconformal algebra is not exactly what we
need to construct the Wakimoto free fields representation of the
$\ssl$ algebra (see Appendix A for details). Indeed, the scalar fields
don't appear.  Nevertheless from \eqref{fssl} it follows that we can
write
\begin{align}\label{qn4c}
  L_n&=\oint_{C_0}dz\,-\gamma^{n+1}J^- -\frac{n+1}{\sqrt
    2}\gamma^n(\xi j^-+\txi j^{\mm})\,(z) \quad,\quad T_n^-=\sqrt
  2\oint_{C_0}dz\,\gamma^n\txi j^-\,(z)\quad, \notag
  \\
  T_n^0&=\oint_{C_0}dz\,\sqrt 2\gamma^n(\xi j^--\txi j^{\mm} )-
  \gamma^n\xi\txi J^-\,(z) \quad,\quad\quad\quad\,\,\, T_n^+=\sqrt
  2\oint_{C_0}dz\,\gamma^n\xi j^{\mm}\,(z)
  \quad, \\
  G^+_r&=2i\oint_{C_0}dz\,\gamma^{r+\oh }(j^{\mm}- \frac{1}{2\sqrt
    2}\xi J^-)\,(z) \quad,\quad G^{\pp}_r=i\sqrt2\oint_{C_0}dz\,
  \gamma^{r+\oh }\xi J^- +\sqrt 2(r+\oh )\xi\txi\gamma^{r-\oh
    }j^{\mm}\,(z)
  \,\,,\notag\\
  G^{\mm}_r&=2i\oint_{C_0}dz\,\gamma^{r+\oh }(j^- -\frac{1}{2\sqrt
    2}\txi J^-)\,(z) \quad,\quad G^-_r=i\sqrt2
  \oint_{C_0}dz\,\gamma^{r+\oh } \txi J^-+\sqrt 2(r+\oh
  )\txi\xi\gamma^{r-\oh }j^-\,(z) \,\,,\notag
\end{align}

Finally, let us discuss a relation with the result of Ito \cite{I}. To
do this we must recall a differential operator realization of the
centerless $N=2$ superconformal algebra. This time it is given by
\begin{equation}\label{n2-4}
  \begin{split}
    L_n&=-\gamma^{n+1}\pd_\gamma-\frac{n+1}{2}\gamma^n(\xi\pd_\xi+\txi\pd\txi
    ) \quad,\quad G_r^+=i\sqrt2\Bigl(a\gamma^{r+\oh
      }\pd_{\txi}+b\gamma^{r+\oh }\xi\pd_{\gamma}+b (r+\oh
    )\xi\txi\gamma^{r-\oh }\pd_{\txi }\Bigr)
    \,\,,\quad \\
    T_n&=\gamma^n(\xi\pd_\xi-\txi\pd\txi ) \quad,\quad
    G_r^-=i\sqrt2\Bigl(c\gamma^{r+\oh }\txi\pd_{\gamma}+ c(r+\oh
    )\txi\xi\gamma^{r-\oh }\pd_{\xi }+ d\gamma^{r+\oh }\pd_{\xi
      }\Bigr) \quad,\quad ac+bd=\oh\,\,.
  \end{split}
\end{equation}
The analysis proceeds as above and leads to the $N=2$ superconformal
algebra with an arbitrary central charge. For a special case we
recover the result by Ito \cite{I}. Moreover, having the $N=2$
superconformal algebra, we can also get the $N=1$ algebra as it has
been done in the previous subsection.


\section{Detailed Examination of Space-Time Symmetries}
\renewcommand{\theequation}{3.\arabic{equation}}
\setcounter{equation}{0} After our rather formal discussion of the
construction of space-time symmetries and their relation to the affine
Lie superalgebras, let us now look more specifically at their relation
to string theory. The simplest example to consider is the bosonic part
of strings on $AdS_3$, i.e. the Virasoro algebra in space-time. This
will be the subject of subsection 3.1.  Later in subsection 3.2 we
will generalize our analysis to include fermions and concentrate on
the properties of the $N=4$ superconformal algebra in space-time.

\subsection{The Virasoro case}

{\it 3.1.1 Reparametrization invariant vertex operators or more?} In
general, the string vertex operators creating physical states are the
primary conformal fields. Moreover, the reparametrization (conformal)
invariance of the theory dictates that the vertices have dimension one
in both $z$ and $\bz$ so that their integral over $dzd\bz$ is
invariant (see e.g., \cite{FMS}). We now wish to apply this to vertex
operators \eqref{Vso}. With the aid of eq.\eqref{stso}, it can be
shown that
\begin{equation*}
  -\Bigl(1+2\frac{f_0}{\alpha_+}(n+1)\Bigr)\gamma^{n+1}J^-+
  2\frac{f_0}{\alpha_+}(n+1)\gamma^nJ^0(z)
\end{equation*}
is the primary conformal field of dimension one only for
$f_0=-\oh\alpha_+$.  It is exactly what was found by Giveon and
co-workers \cite{GKS}.

One of the novelties that appears due to two-dimensional space-time is
that it is possible to give another interpretation. Defining the
Virasoro generators on the world-sheet by the standard way
\begin{equation*}
  {\cal L}_n=\oint_{C_0}dz\,z^{n+1}\Bigl(
  \beta\pd\gamma -\oh\pd\varphi\pd\varphi
  -\frac{1}{\alpha_+}\pd^2\varphi\Bigr)(z)
\end{equation*}
the stringy definition of the vertex operator \eqref{Vso} can be
represented by
\begin{equation}
  [{\cal L}_n,L_m]=0\quad,
\end{equation}
i.e. there are two copies of the Virasoro algebra living on the
world-sheet and in space-time, respectively and these algebras
commute. One of the reasons why this formulation sounds interesting is
that the world-sheet and space-time appear on the equal footing,
independently of which two-dimensional space (world-sheet or
space-time) is primary.

Writing down the stringy vertex operator requires that the scalar
field $\varphi$ has a non-zero background charge, i.e. its central
charge obeys $c\not= 1$. Otherwise the integrand is not a primary
world-sheet conformal field of dimension one. Indeed, what happened is
the anomaly cancelation between $\beta\gamma$ and $\pd\varphi$. So it
is clear that the construction fails for $c=1$. In order to understand
what to do, there is a useful piece of background information we
should review.  First, let us recall that the $\su$ affine Lie algebra
at the level $k=1$ admits the following representation
\begin{equation}
  J^0(z)=\frac{i}{\sqrt 2}\pd\varphi (z)\quad,\quad
  J^{\pm}(z)=\ep^{\pm i\sqrt 2\varphi(z)}\quad.
\end{equation}
We then take the space-time $\su$ generators to be \cite{GKS}
\begin{equation}\label{su2}
  T_n^a=\oint_{C_0} dz\,\gamma^n J^a(z)\quad.
\end{equation}
They obey the commutation relations of $\su$ with $k=p$. So there is a
symmetry enhancement in space-time induced by a proper enhancement on
the world-sheet at $c=1$. However, the main moral of this story is
that the central charge of affine algebras may have a topological
origin! It seems interesting to develop such approach to affine
algebras elsewhere.

{\it 3.1.2 Bosonization or conformal transformation in space-time.}
The construction of section 2 has one drawback: it requires rational
powers of the free field $\gamma(z)$ to describe the superconformal
algebra sectors other than the Neveu-Schwarz sector. Though examples
of fractional calculus for the $(\beta,\gamma)$ system are known in
the literature (see, e.g., \cite{rp} and references therein), it is
more advantageous to use the bosonization procedure \cite{FMS}. The
reason for this is that a special value for space-time dimension, i.e.
$d=2$, allows us to give a geometric interpretation of bosonization.
The point of view we propose is the following. For a given $\gamma(z)$
which classically defines the embedding of the world-sheet into the
space-time, we can take $\gamma(z)={\text e}^{i\sif (z)-\sis (z)}$
that is nothing but a conformal transformation to the cylindrical
coordinates for space-time. On the other hand, within the quantized
$\beta\gamma$ system such representation of $\gamma(z)$ is called as
bosonization. Thus, we can interpret the bosonization procedure as
the quantum version of the conformal transformation to the cylindrical
coordinates in space-time.

At this point, it only remains to illustrate the use of the
cylindrical coordinates in the problem at hand. The general recipe of
$\cite{FMS}$ tells us to set
\begin{equation}\label{bos}
  \gamma(z)=\ep^{i\sif (z)-\sis (z)}\quad,\quad
  \beta(z)=i\pd\sif\,\ep^{-i\sif (z)+\sis (z)}
  \quad,
\end{equation}
where $\sif ,\,\,\sis$ are the scalar fields with proper background
charges and two-point functions normalized as
$\langle\sigma^i(z_1)\sigma^j(z_2)\rangle =-\delta^{ij}\ln\zo$.

Combining all this, Virasoro generators \eqref{qV2} turn out to be
\begin{equation}\label{qVb}
  L_n=\oint_{C_0} dz\,\ep^{n(i\sif-\sis )}
  \Bigl(-(n+1)\pd\sis +in\pd\sif +f(n)\pd\varphi\Bigr)(z)
  \quad.
\end{equation}
A simple check shows that $L_n$ obey \eqref{Vir} with $f(n)=(n+1)f_0$,
$f_0=\text{const}$ and $c=12(1+f_0^2)p$, i.e. the result of subsection
2.1 is recovered.

As regards the operator responsible for the topological charge
\eqref{top}, it is given by
\begin{equation}\label{qtop}
  \hat p=\oint_{C_0} dz\,i\pd\sif -\pd\sis (z)
  \quad.
\end{equation}
An important remark is that $\hat p$ commutes with $\beta(z)$ and
$\gamma(z)$, i.e. with all the $\beta$ and $\gamma$ oscillators. In
fact, this operator is equal to $-Q_{\text B}$, so the number $-p$ is
called the ``Bose-sea level'' of the representation of the $(\beta
,\gamma)$ algebra \cite{V2}.

{\it 3.1.3 Sample calculations of space-time correlators.} Let us now
move towards the world-sheet formulation of perturbative ``string''
theory. In general, physical theory is called string theory if it is
formulated by an underlying world-sheet structure.  What is best
understood presently is the perturbative string theory in
ten-dimensional Minkowski space.  In this case, the space-time
scattering amplitudes in the momentum representation are given by the
functional integral over all two-dimensional matter fields together
with all the geometries of a two-dimensional surface with insertions
of the appropriate vertex operators. It seems natural from the
universality principle to follow the same ansatz. Then it immediately
comes to mind to define space-time correlators in the problem in hand.
A possible way to do this is\footnote{We restrict ourselves to the
  spherical topology, i.e.  tree-level amplitudes. Moreover, we drop
  other matter fields as well as the ghosts.}
\begin{equation}\label{cor-s}
  \langle\,\prod_i O_{n_i}^{h_i}\,\rangle_{st}=
  \bigg\vert\int [d\beta\,d\gamma ]_p\,
\text{exp}(-S_0[\beta,\gamma])\bigg\vert^2
  \int [d\varphi ]\,\text{exp}(-S_0[\varphi])\,
  \text{exp}(\mu S_{\text{int}})\,
  \prod_i\int d^2z\,V_{n_i}^{h_i}(z,\bz )
  \quad.
\end{equation}
Here $S_0[\beta,\gamma]$ and $S_0[\varphi]$ are the standard actions
of the free fields for the Wakimoto representation of $\so$ \cite{W}.
These actions are perturbed by $S_{\text{int}}=\int
d^2z\,\beta\bar\beta\, \text{e}^{-\frac{2}{\alpha_+}\varphi}$ with the
coupling constant $\mu$.  $[d\beta\,d\gamma ]_p$ means that the
$(\beta,\gamma )$ system has the ``Bose-sea level'' $-p$. As to $n_i$
and $h_i$, they are space-time Laurent modes and conformal dimensions,
respectively. We define the measures of the path integrals so as to
have the two-point functions $\langle\varphi(z_1)\varphi(z_2)\rangle
=-\ln\zo ,\,\,\langle\beta(z_1) \gamma(z_2)\rangle_q
=(\frac{z_1}{z_2})^q\frac{1}{\zo}$, where $q$ is the ``Bose-sea
level''.

Before we give examples of computations, we first want to make one
important remark. The point is that the left-hand side of
ansatz~\eqref{cor-s} is a correlator of a proper 2d conformal
(superconformal) theory while a similar ansatz for string theory on
ten-dimensional Minkowski space gives scattering amplitudes of
physical states. Thus what is proposed is the stringy representation
for conformal blocks of 2d conformal field theories.

In order to see how the ansatz \eqref{cor-s} works we will focus on
the correlators of $L_n$. These are, of course, rather simple.
Nevertheless, they contain some technical subtleties.  In fact, modulo
integrations over $z_i$ the right-hand side is a correlator of the
free fields involved in the Wakimoto representation of the $SL(2)$
Wess-Zumino-Witten model. At present there are many ways to compute
such correlators \cite{AF}. The most famous ones are the so-called
Feigin-Fuchs and Dotsenko-Fateev representations. Since we do not know
any conjugate representation of $L_n$ given by \eqref{qV2} with
$f_0=-\oh\alpha_+$, it is natural to make use of the Feigin-Fuchs
representation. For such representation one has a trivial balance of
charges for the free fields
\begin{equation}
  \#\beta=\#\gamma
  \quad,\quad
  \#\alpha_i=0
  \quad.
\end{equation}
Here $\alpha_i$ is given by $V_{n_i}^{h_i}(z)\sim
\text{e}^{\alpha_i\varphi (z)}$.  The price for this simple balance is
an additional number of screening operators that makes integral
representation of conformal blocks more involved than it may be in the
Dotsenko-Fateev representation with a charge asymmetry due to
background charges.  Thus, we have the one-point function
\begin{equation}
  \langle\,L_n\,\rangle_{st}=\oint_{C_0}dz\,
  \bigg\langle\,\text{exp}(\mu\oint d\zp \,\beta\,
  \text{e}^{-\frac{2}{\alpha_+}\varphi (\zp )})
  \Bigl(-\gamma^{n+1}\beta-\oh(n+1)\alpha_+\gamma^n\pd\varphi(z)\Bigr)\,
  \bigg\rangle_{-p}
  \quad.
\end{equation}
Here $-p$ means the ``Bose-sea level'' of the representation of the
$(\beta ,\gamma)$ algebra. Now the balance of charges gives $n=0$ and
the leading term in $\mu$, i.e. $\mu^0$. Thus, the correlator becomes
\begin{equation}
  \langle\,L_n\,\rangle_{st}=-\delta_{n,0}\oint_{C_0}dz\,
  \bigg\langle\,
  \gamma\beta (z)
  \bigg\rangle_{-p}
  \quad.
\end{equation}
The correlator $\langle\,\gamma\beta\rangle_{-p}$ is zero because of
the normal ordering of $\gamma\beta\,\,$\footnote{We use the
  definition $:\beta\gamma(z):=\lim_{z\rightarrow
    w}\big(\beta(w)\gamma(z)-
  \big(\frac{w}{z}\big)^q\frac{1}{w-z}\big)$. Alternatively, using a
  minimal substraction $:\beta\gamma(z):=\lim_{z\rightarrow
    w}\big(\beta(w)\gamma(z) -\frac{1}{w-z}\big)$ leads to
  $\langle\,L_n\,\rangle_{st}=p\,\delta_{n,0}$, i.e. a space-time
  conformal theory has a non-trivial vacuum. However, the
  central charge is the same in both cases.}.  As a result, we have
\begin{equation}
  \langle\,L_n\,\rangle_{st}=0
  \quad.
\end{equation}
This means that a vacuum associated to a space-time conformal theory
looks like the standard SL(2) invariant one.

Next, let us go on to look at the two-point function
\begin{equation}\label{ex2}
  \langle\,L_n L_m\,\rangle_{st}=\prod_{i=1}^2\oint_{C_i}dz_i\,
  \bigg\langle\,\text{exp}(\mu\oint d\zp \,\beta\,
  \text{e}^{-\frac{2}{\alpha_+}\varphi (\zp )})\prod_{i=1}^2
  \Bigl(-\gamma^{n+1}\beta-\oh(n+1)\alpha_+\gamma^n\pd\varphi(z_i)\Bigr)\,
  \bigg\rangle_{-p}
  \quad.
\end{equation} 
The integration contours for $z_1,\,z_2$ are shown in Fig.1.

%
\vspace{.3cm}
\begin{figure}[ht]
  \begin{center}
    \includegraphics{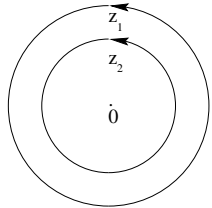}
    \caption{Contours used in the definition of the 
      two-point correlation function.}
  \end{center}
\end{figure}

\vspace{-0.3cm} It is easy to see that the balance of charges and the
integration over $\varphi$ yield
\begin{equation}
  \langle\,L_n L_m\,\rangle_{st}=\delta_{n+m,0}
\prod_{i=1}^2\oint_{C_i}dz_i\,
  \bigg\langle\,\gamma^{n+1}\beta (z_1)\gamma^{m+1}\beta (z_2)-
  \oq (n+1)(m+1)\alpha_+^2\gamma^n(z_1)\gamma^m(z_2)\frac{1}{\zt }
  \,\bigg\rangle_{-p}
  \quad.
\end{equation} 
It is also not difficult to integrate $\beta$ away. This can be done
by simply using the two-point function of $(\beta,\gamma)$. Thus the
correlator becomes
\begin{equation}
  \langle\,L_n L_m\,\rangle_{st}=-(n+1)(m+1)(1+\oq\alpha_+^2)\,\delta_{n+m,0}
  \prod_{i=1}^2\oint_{C_i}dz_i\,\frac{1}{\zt }
  \bigg\langle\,\gamma^n(z_1)\gamma^{-n} (z_2)
  \,\bigg\rangle_{-p}
  \quad.
\end{equation} 
The last correlator is computed in Appendix B. So we may write
\begin{equation}
  \langle\,L_n L_m\,\rangle_{st}=-\frac{k}{2}(n+1)(m+1)\,\delta_{n+m,0}
  \prod_{i=1}^2\oint_{C_i}dz_i\,\frac{1}{\zt }\Big(\frac{z_1}{z_2}\Big)^{pn}
  \quad.
\end{equation} 
In above we also set $\alpha_+=\sqrt{2(k-2)}$.

Now we come to the analysis of the integrals. Obviously, the result
depends on $p$:

(i) For $p=0$, the integral over $z_2$ is equal to zero because the
corresponding contour may be contracted to a point. As a result, the
correlator vanishes.

(ii) For $p>0$ and $n>0$, the contour $C_2$ with $z_1$ fixed can be
deformed as shown in Fig.2, so the integral is given by the residue of
the integrand. As a result, we have
\begin{equation}
  \langle\,L_n L_m\,\rangle_{st}=\oh kp(n^3-n)\,\delta_{n+m,0}
  \quad.
\end{equation} 
Thus the central charge is given by $c=6kp$, which is in agreement
with the result of section 2.

%
\vspace{.4cm}
\begin{figure}[ht]
  \begin{center}
    \includegraphics{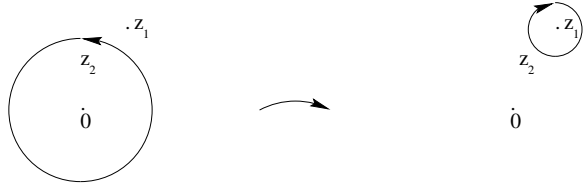}
    \caption{Deformation of the contour $C_2$.}
  \end{center}
\end{figure}

\vspace{-0.3cm} 
For $p>0$ and $n<0$, the integral over $z_2$ equals
zero because the contour $C_2$ may be contracted to a point, so the
correlator vanishes.

(iii) For $p<0$, the analysis proceeds as in the above. Indeed, this
is the same as (ii) with $\gamma\rightarrow\gamma^{-1}$.

There are two main conclusions:

(a) The proposed ansatz \eqref{cor-s} together with the corresponding
normal ordering prescription reproduces the standard SL(2) invariant
vacuum of space-time conformal field theory, namely $L_n\vert
0\rangle_{st} =0\,\,\text{for}\,\,n\geq -1$. Moreover, it also
provides the conjugate vacuum by (iii) with ${}_{st}\langle 0\vert
L_n=0\,\,\text{for}\,\,n\leq 1$. Eigenvalues of the topological charge
$\hat p$ have to be restricted to the fundamental region $p\geq 0$.

(b) There are no corrections to the central charge due to the
perturbation term $S_{\text{int}}$. However, it is expected that this
term is relevant for the structure of space-time conformal theory. One
example how it can be realized is known. We mean the deformation of
the chiral algebra ($w_\infty$) of 2d gravity coupled to $c\leq 1$
matter \cite{cc}. Recall that such deformation appears due to the
expansion over the cosmological term in the Liouville action which is
nothing but the screening operator for the Virasoro algebra. So, the
only difference between these cases is the symmetry algebra on the
world-sheet: Virasoro for 2d gravity and $\so$ in the case of
interest.

It is a straightforward matter of working out the multi-point
functions $\langle \prod_i L_{n_i}\rangle_{st}$ and then deriving the
Ward identities.  In fact, the latter requires repeating the procedure
of \S 2.

Let us conclude this subsection with a remark.

The computations of section 2 are based on zero ``Bose-sea level'' for
the $(\beta,\gamma)$ system. On the other hand, we have used a
non-zero level in this section. So the problem is to see what happens
if we use a non-zero level for the computations of section 2, i.e.
we should check that the Virasoro generators given by \eqref{qV2} are
correct. As in section 2, a simple algebra shows that such defined
$L_n$ obey the Virasoro commutation relations only for $f(n)=(n+1)f_0$
with $f_0=\text{const}$. Thus the definition of $L_n$ does not depend
on the ``Bose-sea level'' of the $(\beta,\gamma)$ system.
Alternatively, this can be understood as follows. By using
bosonization of the $(\beta,\gamma)$ system, we represent the
q-vacuum via the vertex operator at $z=0$. It is well known that a
vertex operator sitting at the origin does not affect the commutators
of operators defined via contour integrals surrounding $0$.

{\it 3.1.4 $\varphi\rightarrow\infty$ limit.} According to the
Maldacena conjecture \cite{Mal}, string theory on $AdS_3$ times a
compact space is dual to a CFT whose world-sheet is the boundary of
$AdS_3$. First let us clarify what the boundary is. Starting with the
metric on $AdS_3\,$\footnote{We use the Euclidean version with radius
  $l$.}
\begin{equation}
  ds^2=\frac{l^2}{r^2}(dr^2+d\gamma d\bar\gamma )\quad,
\end{equation}
the boundary consists of a copy of $\text{\bf R}^2$ at $r\rightarrow
0$, which in the $(\varphi ,\gamma ,\bar\gamma)$ coordinates
corresponds to $\varphi\rightarrow +\infty$, together with a single
point at $r\rightarrow\infty$, or $\varphi\rightarrow -\infty$. Thus,
in this representation, the boundary of $AdS_3$ is obtained by adding
a point at infinity to $\text{\bf R}^2$, which is nothing but a sphere
$\text{\bf S}^2$.

One basic question we should ask is how to implement ``the boundary
conditions'' in the path integral \eqref{cor-s}. To answer this
question, let us recall that the CFT/AdS correspondence is holographic
in sense of \cite{H,LS}.  This means that a quantum theory with
gravity must be describable by a boundary theory. It seems natural
therefore to require that stringy vertex operators are independent of
$\varphi$. As it is shown in section 2 we cannot define the
generators of Virasoro and $N=2$ algebras in a $\varphi$-independent
way but we can do so in the case of $N=4$ when $\varphi$ drops
out.\footnote{We do not know any different $\varphi$-independent
  representation for stringy vertex operators that has been discussed
  in the literature on this subject [6, 8-13].} Thus the Holographic
principle is in harmony with the Maldacena conjecture that one should
get exactly $N=4$ on the boundary!

{\it 3.1.5 Central charge multiplication.} Given the Virasoro algebra
with central charge $c$, it is a simple thing to build the algebra
with a central charge divisible by $c$, i.e. $\tilde c=\tp c$, where
$\tp$ is a positive integer. The actual relation is
\begin{equation}\label{tVir}
  \tilde L_n=\frac{1}{\tp }L_{n\tp }+
  \frac{c\tp }{24}\Bigl(1-\frac{1}{\tp^2}\Bigr)\delta_{n,0}
  \quad.
\end{equation}
Then we can write $\tilde L_n$ in terms of $\beta ,\gamma ,\varphi$
as
\begin{equation}
  \tilde L_n=\frac{1}{\tp }\oint_{C_0} dz\,-\gamma^{n\tp +1}\beta
  +f(n\tp )\gamma^{n\tp }\pd\varphi(z)\,\,
  +\frac{c\tp }{24}\Bigl(1-\frac{1}{\tp^2}\Bigr)\delta_{n,0}
  \quad.
\end{equation}

This mechanism actually has the following interesting interpretation.
What we had before was a map $\gamma$:$\,U_0\rightarrow U_\gamma$,
i.e. a local embedding of the world-sheet into space-time. Now let us
define a covering map of $U_\gamma$ on $\text{\bf CP}^1$ namely,
$\tga$:$\,U_\gamma\rightarrow U_{\tga } $(a patch of $\text{\bf
  CP}^1$). If the covering map takes the form $\tga =\gamma^{\tp }$,
then eq.\eqref{tVir} is nothing but the consequence of the
transformation law of the stress-energy tensor under the analytic
change of coordinates (see \cite{VK}). So we can rewrite $\tilde L_n$
as
\begin{equation}
  \tilde L_n=\frac{1}{\tp }\oint_{C_0} dz\,
  -\tga^{n+\frac{1}{\tp }}\beta
  +f(n\tp )\tga^n\pd\varphi(z)\,\,
  +\frac{c\tp }{24}\Bigl(1-\frac{1}{\tp^2}\Bigr)\delta_{n,0}
  \quad.
\end{equation}
Thus from the geometrical point of view the mechanism gives a short
sequence of maps
\begin{equation*}
  U_0\rightarrow U_\gamma\rightarrow U_{\tga }
\end{equation*}
that is nothing but a particular case of the Green's proposal
``world-sheets for world-sheets''! Of course we can immediately
generalize the above mechanism for superalgebras. We will do this in
the next subsection.

\subsection{Superalgebras}
{\it 3.2.1 More sample calculations of space-time correlators.} First
we want to see the world-sheet conformal invariance of
vertices~\eqref{qn4}. It immediately follows from the fact that their
integrands have dimension one in $z$. In other words, defining the
Virasoro generators on the world-sheet as
\begin{equation*}
  {\cal L}_n=\oint_{C_0}dz\,z^{n+1}\Bigl(
  \beta\pd\gamma +\pd\xi\eta+\pd\txi\te +\dots\Bigr)(z)
  \quad,
\end{equation*}
where the dots mean extra contributions from $\varphi$ and
$\tilde\varphi$, we simply get that ${\cal L}_n$ commutes with the
generators \eqref{qn4}. Thus we have the stringy representation for
the space-time $N=4$ superconformal algebra.

Now let us give some examples of calculations of space-time
correlators. In doing so, for simplicity of exposition we restrict
ourselves to the spherical topology and omit fields other than those
used to define the $N=4$ generators \eqref{qn4}. Note that according
to our discussion of the $\varphi\rightarrow\infty$ limit and
Holographic principle in subsection 3.1.4, $\varphi$ must be omitted.
Now ansatz \eqref{cor-s} is replaced by
\begin{equation}\label{cor-b}
  \langle\,\prod_i O_{n_i}^{h_i\,j_i}\,\rangle_{st}=
  \bigg\vert\iiint [d\beta\,d\gamma ]_p\,[d\eta\,d\xi]\,[d\te \,d\txi ]
  \,\text{exp}(-S_0)\bigg\vert^2
  \text{exp}(\mu S_{\text{int}})\,
  \prod_i\int d^2z\,V_{n_i}^{h_i\,j_i}(z,\bz )
  \quad.
\end{equation}
Here $S_0$ is the sum of the standard actions for the first order
systems. $[d\beta\,d\gamma ]_p$ means that the $(\beta,\gamma )$
system has the ``Bose-sea level'' $-p$.  As regards $n_i$, $h_i$ and
$j_i$, they are space-time Laurent modes, conformal dimensions and
SU(2)-spins, respectively.  The measures in the path integrals are
defined to have the two-point functions
$\langle\beta(z_1)\gamma(z_2)\rangle_q
=(\frac{z_1}{z_2})^q\frac{1}{\zo},\,\,\langle\eta(z_1)\xi(z_2)\rangle=
\langle\te (z_1)\txi(z_2)\rangle=\frac{1}{\zo}$, where $q$ is the
``Bose-sea level''. Of course, the free action $S_0$ may be perturbed
as it was done earlier. One basic question we should now ask is what
$S_{\text{int}}$ is. A possible solution is to take the screening
operators of $\ssl$ as the perturbation.  Fortunately, what saves the
day is that the explicit form of $S_{\text{int}}$ is irrelevant for
what follows.

We will now carry out computations for the $N=4$ superalgebra using
the above ansatz. Let us again choose the representation a la
Feigin-Fuchs with the following balance of charges
\begin{equation}\label{bal}
  \#\beta=\#\gamma
  \quad,\quad
  \#\eta=\#\xi
  \quad,\quad
  \#\te=\#\txi
  \quad.
\end{equation}

The one-point functions can be evaluated to give
\begin{equation}
  \langle\,L_n\,\rangle_{st}=\langle\,T_n^a\,\rangle_{st}
  =\langle\,G_r^\alpha\,\rangle_{st}=0
  \quad,
\end{equation}
where, just for simplicity, $n\in\text{\bf Z}$ and $r\in\text{\bf
  Z}+\oh$.  We use the same definition for normal ordering as in
subsection 3.1.3. So, the string vertices defined in this way lead to
the standard NS vacuum of the space-time CFT.

Next, let us go on to look at some two-point functions.

The first example is $\langle\,L_n L_m\,\rangle_{st}$. By using the
balance \eqref{bal} and integrating over the free fields, we can
write it as
\begin{equation}
  \langle\,L_n L_m\,\rangle_{st}=(n^2-1)\,\delta_{n+m,0}
  \prod_{i=1}^2\oint_{C_i}dz_i\,\frac{1}{\zt }\Big(\frac{z_1}{z_2}\Big)^{pn}
  \quad.
\end{equation} 
To analyze the integrals we proceed as in subsection 3.1.3. For
$p>0\,,\,\,n>0$ the result is
\begin{equation}
  \langle\,L_n L_m\,\rangle_{st}=\frac{p}{2}(n^3-n)\,\delta_{n+m,0}
  \quad,
\end{equation} 
that gives the central charge $c=6p$.

At this point it is natural to make a double check of the above result
for $c$.  For this purpose, let us compute $\langle\,T^0_n
T^0_m\,\rangle_{st}$. All computations proceed as before and result in
\begin{equation}
  \langle\,T^0_n T^0_m\,\rangle_{st}
  =\frac{p}{2}n\,\delta_{n+m,0}
  \quad,
\end{equation} 
where $n,\,\,p$ are positive integers. So, $c$ is again equal to $6p$.
If instead of $\langle\,T^0_n T^0_m\,\rangle_{st}$ the two-point
function is $\langle\,G^+_r G^-_s\,\rangle_{st}$, then a simple
algebra shows that for $p>0\,,\,\,r>\oh$
\begin{equation}
  \langle\,G^+_r G^-_s\,\rangle_{st}
  =2p(r^2-\oq )\,\delta_{r+s,0}
  \quad,
\end{equation} 
i.e. $c=6p$, which is the expected result.

It is straightforward to find other two-point functions as well as
n-point ones.

{\it 3.2.2 The $N=4$ case, revisited.} Up to now our discussion of the
$N=4$ superconformal algebra has been somewhat incomplete since the
set of the free fields used to build generators \eqref{qn4} is not
exactly what one needs to construct the Wakimoto free field
representation of the $\ssl$ algebra. So we can say that the
underlying world-sheet algebra may be different. Let us now fill this
gap in our discussion. In doing so, we need to note just one point:
for the particular set of the free fields that we are discussing here,
we can define the following representation of the affine Lie
superalgebra $\ssl$ at the level $k=1\,\,$\footnote{Note that the thus
  defined $\ssl$ has $\hat{su}(2)$ with $k=1$ as a subalgebra, which
  makes it different from what is represented in Appendix A.}
\begin{alignat}{4}
  J^+(z) & =\xi\te (z)\,, & \quad J^-(z) & =\txi\eta (z)\,, & \quad
  J^0(z) & =\oh (\xi\eta -\txi\te )(z)\,, & \quad J^3(z) & =
  \beta\gamma+\oh (\xi\eta +\txi\te )(z)\,,\notag
  \\
  j^+(z) & = \frac{1}{\sqrt 2}\gamma\te (z)\,, & \quad j^{\pp }(z) & =
  \frac{1}{\sqrt 2}\xi\beta (z)\,, & \quad j^-(z) & = \frac{1}{\sqrt
    2}\txi \beta\,, & \quad j^{\mm}(z) & =\frac{1}{\sqrt 2}\gamma\eta
  (z) \quad.
\end{alignat}
After this is done, the generators of $N=4$ are immediately written as
\begin{alignat}{2}\label{n4-ssl}
  L_n & =-\oint_{C_0}dz\, \gamma^nJ^3+\frac{n}{\sqrt
    2}\gamma^{n-1}(\xi j^{\mm }+\txi j^+)\,(z) \quad, & \quad T_n^a &
  =\oint_{C_0}dz\, \gamma^nJ^a\,(z)
  \quad, \notag\\
  G^{\pp}_r & =2i\oint_{C_0}dz\, \gamma^{r+\oh }j^{\pp
    }-\frac{1}{\sqrt 2}\gamma^{r-\oh }\txi J^+\,(z) \quad, & \quad\,\,
  G^+_r & =2i\oint_{C_0}dz\, \gamma^{r-\oh }j^+\,(z)
  \quad,\\
  G^-_r & =2i\oint_{C_0}dz\, \gamma^{r+\oh }j^--\frac{1}{\sqrt
    2}\gamma^{r-\oh }\xi J^-\,(z) \quad, & \quad\,\, G^{\mm}_r &
  =2i\oint_{C_0}dz\, \gamma^{r-\oh }j^{\mm }\,(z) \quad.\notag
\end{alignat}
Thus $\ssl$ is indeed the underlying world-sheet algebra.

To complete the story, perhaps we should point out that \eqref{n4-ssl}
can be regarded as one of the generalizations of the construction
\eqref{su2}.

{\it 3.2.3 NS fivebranes.} According to the AdS/CFT correspondence, a
CFT whose world-sheet is the boundary of $AdS_3$ corresponds to the IR
limit of the dynamics of parallel $Q_1$ $D_1$-branes and $Q_5$
$D_5$-branes that is mapped under S-duality into $Q_1$ fundamental
strings and $Q_5$ NS5-branes.  The central charge of this boundary CFT
is given by $6Q_1Q_5$. One consequence of the construction of strings
on $AdS_3$ is that $p$ measures the number of fundamental strings,
i.e. $p=Q_1$. As to $Q_5$, it is identified with the level of the
world-sheet algebra $\so$ \cite{GKS}.  The latter does not seem
natural from either the ``world-sheets for world-sheets'' standpoint
or from the topological origin of central charges, so it is reasonable
to ask how the situation can be improved.  Actually, as we have noted
at the end of subsection 3.1, there is a nice way to do this. The
intuitive idea is that $c\rightarrow c\tp $. So, if $c=6p$ then
$\tilde c=6p\tp$ and $\tp$ is expected to be the number of NS5-branes.
It is easy to see that the $c$ and $\tilde c$ generators of the $N=4$
superconformal algebra are related by
\begin{equation}
  \tilde L_n=\frac{1}{\tp }L_{n\tp }+
  \frac{c\tp }{24}\Bigl(1-\frac{1}{\tp^2}\Bigr)\delta_{n,0}
  \quad,\quad
  \tilde T^a_n=T^a_{n\tp }
  \quad,\quad
  \tilde G^{\alpha}_r=\frac{1}{\sqrt{\tp} }G^{\alpha}_{r\tp }
  \quad.
\end{equation}
Thus in terms of the free fields the $\tilde c$ generators are
\begin{equation}\label{tn4}
  \begin{split}
    \tilde L_n&=\frac{1}{\tp }\oint_{C_0}dz\, -\gamma^{n\tp +1}\beta
    -\frac{n\tp +1}{2}\gamma^{n\tp }(\xi\eta+\txi\te )\,(z)\,\,+
    \frac{c\tp }{24}\Bigl(1-\frac{1}{\tp^2}\Bigr)\delta_{n,0}
    \quad,\quad \\
    \tilde T_n^0&=\oh\oint_{C_0}dz\,\gamma^{n\tp }(\xi\eta-\txi\te
    )\,(z) \quad,\quad \tilde T_n^-=\oint_{C_0}dz\,\gamma^{n\tp
      }\txi\eta\,(z) \quad,\quad \tilde
    T_n^+=\oint_{C_0}dz\,\gamma^{n\tp }\xi\te\,(z)
    \quad,\\
    \tilde G^+_r&=i\sqrt{\frac{2}{\tp }} \oint_{C_0}dz\,\gamma^{r\tp
      +\oh }\te\,(z) \quad,\quad\quad \tilde
    G^{\pp}_r=i\sqrt{\frac{2}{\tp }} \oint_{C_0}dz\, \gamma^{r\tp +\oh
      }\xi\beta +(r\tp +\oh )\xi\txi\gamma^{r\tp -\oh }\te\,(z)
    \quad,\quad \\
    \tilde G^{\mm}_r&=i\sqrt{\frac{2}{\tp }}
    \oint_{C_0}dz\,\gamma^{r\tp +\oh }\eta\,(z) \quad,\quad\quad
    \tilde G^-_r=i\sqrt{\frac{2}{\tp }} \oint_{C_0}dz\,\gamma^{r\tp
      +\oh } \txi\beta+(r\tp +\oh )\txi\xi\gamma^{r\tp -\oh }\eta\,(z)
    \quad.
  \end{split}
\end{equation}
{}From the geometrical point of view what we proposed is again a short
sequence of maps
\begin{equation*}
  U_0\rightarrow U_\gamma\rightarrow U_{\tga }\quad,
\end{equation*}
where $\tga =\gamma^{\tp }$, which allows us to interpret NS5-branes
as strings whose world-sheet is a manifold with the coordinates
$(\gamma,\bar\gamma)$.


\section{Conclusions and Remarks}
\renewcommand{\theequation}{3.\arabic{equation}}
\setcounter{equation}{0}

First let us say a few words about the results.

In this work we have considered the possible application of postulates
(i)-(iii), as these are formulated in \S 1, to superstring theory on
$AdS_3$. Starting with a formal construction which lifts bosonized
affine Lie superalgebras on an arbitrary world-sheet to
two-dimensional space-time we found rather amusing relations between
space-time $N=2,\,4$ superconformal algebras and world-sheet
superalgebras $\os ,\,\,\ssl$. Next following the postulates, we
applied this construction to superstring propagation on $AdS_3$. We
succeeded in describing the boundary $N=4$ algebra in terms of the
``boundary'' coordinates only, as postulate (i) requires, and in
interpreting the central charge of this algebra in both ``primitive''
and topological ways, as postulates (ii)-(iii) require.  We have also
given a prescription for the calculation of space-time correlators. As
an example, we computed the central charge via two-point functions.

Let us conclude by mentioning a few open problems together with
interesting features of our discussion of string on $AdS_3$.

(i) Obviously, the most important open problem is to find the spectrum
of the theory by solving the corresponding BRST problem and establish
unitarity \footnote{In fact, we know what answer should be namely, the
  unitary representations of $N=4$. So the problem is to give their
  stringy representation.}. Unfortunately, it is unknown in general
how to do this.  Most of the approaches have so far been based on the
world-sheet $\hat{sl}(2,R)$ symmetry algebra and its Wakimoto
representation. This may cause new problems because $\hat{sl}(2,R)$
has no unitary representations \cite{DPL}\footnote{Note that it is
  possible to catch unitarity by modifying the Wakimoto representation
  as it was done by Bars \cite{B}.}.

(ii) From the mathematical point of view stringy representations of
superconformal algebras are interesting in their own setting. These
are not presented in the literature, so further work is needed to
develop both the theory of such representations and their applications
to conformal field theory.

(iii) It is interesting to note that $N=4$ is special because it
provides the example where both the Holographic principle and
unitarity\footnote{At least, the central charge turns out to
  correspond to the unitary representations.} are simultaneously
present. Moreover, postulates (ii)-(iii) can be fitted in with this
case. All that provides one more evidence that the AdS/CFT
correspondence should make sense.

(iv) In our discussion of the ``world-sheets for world-sheets'' in
section 3 a short sequence of maps $U_0\rightarrow U_\gamma\rightarrow
U_{\tga }$ requires one functional integration while the original
procedure assumes one integration per map. So, it seems more natural
to call it the simplified ``world-sheets for world-sheets'' rather
than the Green's ``worlds-sheets for world-sheets''.

(v) The construction of the stringy representations of superalgebras
or, strings on $AdS_3$ as a special case, is strongly reminiscent of
the non-critical string theory in sense that both of them have
two-dimensional space-time. However, the difference is in how the
corresponding embedding of the world-sheet is implemented. In the
first case it is done by the first order systems that is natural from
both the geometrical point of view and as regards the manifest
supersymmetry in space-time.  As to the second, it is realized by the
scalar fields.  Nevertheless, these constructions should have
relations.  Some similarities have been noticed in \cite{GKP,M}.
However more work is needed to determine a precise relation.

\vspace{.25cm} {\bf Acknowledgments}

\vspace{.25cm} I heard about the Green's ``world-sheets for
world-sheets'' during my visit to Ecole Normale Sup\'erieure in
1996-1997. I am indebted to J.-L.Gervais for stimulating discussions
of this nice idea. I am also grateful B. Feigin, R. Metsaev and A.
Rosly for useful discussions, and A. Semikhatov for reading the
manuscript. This research was supported in part by the European
Community grant INTAS-OPEN-97-1312.


\appendix
\section{Some affine Lie superalgebras and their free field representations}
\renewcommand{\theequation}{A.\arabic{equation}}
\setcounter{equation}{0}

In this appendix we will briefly recall basic facts on some affine Lie
superalgebras and their free field representations. All superalgebras
have $\hat {sl}(2,R)$ as a subalgebra that makes them different from
what can usually be found in the literature (see, e.g., \cite{W} and
references therein).

\subsection{The $\So$ case}

As a preparation for the discussion of superalgebras in later
subsections, let us begin with the affine Lie algebra $\so$. In the
case of $\So$ the algebra consists of 3 bosonic currents required to
obey the following Operator Product expansions
\begin{equation}\label{sl2}
  J^a(z_1)J^b(z_2)=\frac{k}{2}\frac{g^{ab}}{\zt}+
  \frac{f^{ab}_c}{\zo}J^c(z_2)+O(1)\quad,
\end{equation}
where $k$ is the level, $g^{00}=-1,\,\,g^{+-}=g^{-+}=2,\,\,f^{0+}_+=
f^{-0}_-=1,\,\,f^{-+}_0=2;\,\,a,b,c=0,+,-$.

It is well-known that the Wakimoto free field representation is
described in terms of one free boson $\varphi$ coupled to a background
charge and a first order bosonic $(\beta ,\gamma)$ system of weight
$(1,0)$. The two-point functions of the free fields are normalized as
\begin{equation}
  \langle \varphi(z_1)\varphi(z_2)\rangle=-\ln\zo\quad,\quad
  \langle \beta(z_1)\gamma(z_2)\rangle=\frac{1}{\zo}\quad.
\end{equation}
The currents are given by
\begin{align}\label{so}
  J^-(z)&=\beta(z)\quad,\notag\\
  J^0(z)&=\beta\gamma+\frac{\alpha_+}{2}\pd\varphi(z)\quad,\\
  J^+(z)&=\beta\gamma^2+\alpha_+\gamma\pd\varphi+k\pd\gamma(z)\quad,\notag
\end{align}
with $\alpha_+=\sqrt{2(k-2)}$.

One can use the Sugawara construction, namely
\begin{equation}
  T(z)=\frac{1}{k-2}g_{ab}J^aJ^b(z)\quad,
\end{equation}
to determine the Virasoro algebra with the central charge
$c=\frac{3k}{k-2}$.

Moreover, with the aid of eq.\eqref{so}, $T(z)$ can be written as
\begin{equation}\label{stso}
  T(z)=\beta\pd\gamma -\oh\pd\varphi\pd\varphi
  -\frac{1}{\alpha_+}\pd^2\varphi(z)\quad.
\end{equation}

Finally, let us note that we can write some free fields in terms of
the currents
\begin{equation}\label{fso}
  \beta(z)=J^-(z)\quad,\quad 
  \pd\varphi(z)=\frac{2}{\alpha_+}(J^0-\gamma J^-)(z)\quad.
\end{equation}

\subsection{The $\Os$ case}
The simplest affine superalgebra $\os$ is obtained from the algebra
$\so$ by incorporating fermionic currents $j^\pm(z)$. Thus the OP
expansions \eqref{sl2} are extended to
\begin{align}
  J^a(z_1)J^b(z_2)&=\frac{k}{2}\frac{g^{ab}}{\zt}+
  \frac{f^{ab}_c}{\zo}J^c(z_2)+O(1)\quad,\notag \\
  J^a(z_1)j^\alpha(z_2)&=\frac{(\sigma^a)^\alpha_\beta}{\zo}
  j^\beta(z_2)+
  O(1)\quad, \\
  j^\alpha(z_1)j^\beta(z_2)&
=\frac{k}{2}\frac{\varepsilon^{\alpha\beta}}{\zt}+
  \frac{(\sigma^a)^{\alpha\beta}}{\zo}J_a(z_2)+ O(1)\quad, \notag
\end{align}  
with $\alpha,\,\beta=+,-$; $J_a=g_{ab}J^b$, $(\sigma^a)^{\alpha\beta}
=(\sigma^a)^{\alpha}_\gamma \varepsilon^{\gamma\beta}$, where
$\varepsilon$ is the antisymmetric tensor with
$\varepsilon^{+-}=-\varepsilon^{-+}=1$; and the tensors
$(\sigma^a)^\alpha_\beta$ are in the representation $(\sigma^0)^+_+=
-(\sigma^0)^-_-=\oh\,,\,\,(\sigma^-)^+_-=-(\sigma^+)^-_+=1$.

This time the Wakimoto free field representation is described in terms
of one free boson $\varphi$ coupled to a background charge and a pair
of first order systems $(\beta ,\gamma)$ and $(\eta ,\xi)$ of weights
$(1,0)$. The two-point function of the free fermions is normalized as
\begin{equation}
  \langle \eta(z_1)\xi(z_2)\rangle=\frac{1}{\zo}\quad.
\end{equation}
The currents are given by
\begin{align}\label{os}
  J^-(z)&=\beta(z)\quad,\notag\\
  J^0(z)&=\beta\gamma+\oh\xi\eta+\alpha_+\pd\varphi(z)\quad,\notag\\
  J^+(z)&=\beta\gamma^2+\gamma\xi\eta +2\alpha_+\gamma\pd\varphi
  +k\pd\gamma +\oq (1-k)\xi\pd\xi(z)\quad,\\
  j^-(z)&=\eta -\frac{1}{4}\xi\beta(z)\quad, \notag \\
  j^+(z)&=\gamma\eta -\frac{1}{4}\xi\gamma\beta
  -\oh\alpha_+\xi\pd\varphi + \oq (1-2k)\pd\xi(z)\quad,\notag
\end{align}
where $\alpha_+=\oh\sqrt{2k-3}$.

For $\os$, the Sugawara stress tensor
\begin{equation}
  T(z)=\frac{1}{k-\frac{3}{2}}\Bigl(g_{ab}J^aJ^b+\varepsilon_{\alpha\beta}
  j^\alpha j^\beta\Bigr)(z)\,,\,\, \text{ with }\,\,
  \varepsilon_{\alpha\beta}\varepsilon^{\beta\gamma}=\delta^\gamma_\alpha
\end{equation}
gives the Virasoro algebra with the central charge
$c=\frac{k}{k-\frac{3}{2}}$.

In terms of the free fields $T(z)$ is given by
\begin{equation}
  T(z)=\beta\pd\gamma +\pd\xi\eta -\oh\pd\varphi\pd\varphi
  -\frac{1}{4\alpha_+}\pd^2\varphi(z)\quad.
\end{equation}

Finally, let us note that we can also write some free fields via the
currents as
\begin{equation}\label{fos}
  \beta(z)=J^-(z)\quad,\quad 
  \eta(z)=j^-+\oq\xi J^-(z)\quad,\quad
  \pd\varphi(z)=\frac{1}{\alpha_+}(J^0-\gamma J^--\oh\xi j^-)(z)\quad.
\end{equation}

\subsection{The $\Sl$ case}
The next affine superalgebra to consider is $\ssl$. It is obtained
from from the algebra $\os$ by incorporating fermionic currents
$j^{\dpm}$ as well as one bosonic current $J^3$. We take the defining
OP expansions to be
\begin{align}
  J^a(z_1)J^b(z_2)&=\frac{k}{2}\frac{g^{ab}}{\zt}+
  \frac{f^{ab}_c}{\zo}J^c(z_2)+O(1)\quad,\notag \\
  J^a(z_1)j^\alpha(z_2)&=\frac{(\sigma^a)^\alpha_\beta}{\zo}
  j^\beta(z_2)+
  O(1)\quad, \\
  j^\alpha(z_1)j^\beta(z_2)&
=\frac{k}{2}\frac{\varepsilon^{\alpha\beta}}{\zt}+
  \frac{(\sigma^a)^{\alpha\beta}}{\zo}J_a(z_2)+ O(1)\quad, \notag
\end{align}  
where $g^{33}=1$; $a,b,c=0,\pm,3$; $\alpha,\,\beta=+,-,\pp ,\mm$. This
time $\varepsilon$ is the antisymmetric tensor with
$\varepsilon^{+-}=\varepsilon^{\pp\mm}=1$; and the tensors
$(\sigma^a)^\alpha_\beta$ are in the representation $(\sigma^0)^+_+=
(\sigma^3)^+_+=(\sigma^0)^{\pp}_{\pp}=(\sigma^3)^{\mm}_{\mm}=\oh$,
$(\sigma^0)^-_-=(\sigma^3)^-_-=(\sigma^0)^{\mm}_{\mm}=
(\sigma^3)^{\pp}_{\pp}=-\oh$,
$(\sigma^+)^-_{\pp}=-(\sigma^-)^+_{\mm}=1$.  Moreover,
$(\sigma^a)^\alpha_\beta=(\sigma^a)^\gamma_\lambda
\varepsilon_{\gamma\beta}\varepsilon^{\lambda\alpha}$.

The Wakimoto free field representation is described in terms of a pair
of free bosons $(\varphi, \phi)$ without background charges and a set
of first order systems $(\beta ,\gamma)$, $(\eta ,\xi)$ and $(\te
,\txi)$ of weights $(1,0)$. The two-point functions of the free fields
are normalized as
\begin{equation}
  \langle \phi(z_1)\phi(z_2)\rangle=-\ln\zo\quad,\quad
  \langle \te(z_1)\txi(z_2)\rangle=\frac{1}{\zo}\quad.
\end{equation}

For the bosonized currents, we take
\begin{align}\label{ssl}
  J^-(z)&=\beta(z)\quad,\notag\\
  J^0(z)&=\beta\gamma +\oh(\xi\eta+\txi\te )+\alpha_
+\pd\varphi(z)\quad,\notag\\
  J^3(z)&=\oh\xi\eta -\oh\txi\te +i\alpha_+\pd\phi (z)\quad,\notag \\
  J^+(z)&=\beta\gamma^2 +\gamma(\xi\eta +\txi\te )+k\pd\gamma +
  2\alpha_+\gamma\pd\varphi +\oh (k-1)(\txi\pd\xi+\xi\pd\txi )+
  i\alpha_+\xi\txi\pd\phi(z)\quad,\\
  j^-(z)&=\frac{1}{\sqrt 2}\Bigl(\eta +\oh\txi\beta\Bigr)(z)\quad, \notag \\
  j^+(z)&=\frac{1}{\sqrt 2}\Bigl(-\gamma\te -\oh\xi\gamma\beta
  -\oh\xi\txi\te+\oh (1-2k)\pd\xi-
  \alpha_+\xi (\pd\varphi-i\pd\phi)\Bigr)(z) \quad, \notag \\
  j^{\mm}(z)&
=\frac{1}{\sqrt 2}\Bigl(\te +\oh\xi\beta\Bigr)(z)\quad, \notag \\
  j^{\pp}(z)&=\frac{1}{\sqrt 2}\Bigl(-\gamma\eta -\oh\txi\gamma\beta
  -\oh\txi\xi\eta+ \oh (1-2k)\pd\txi-\alpha_+\txi
  (\pd\varphi+i\pd\phi)\Bigr)(z) \quad,\notag
\end{align}
where $\alpha_+=\sqrt{\oh (k-1)}$.

For $\ssl$, the Sugawara stress tensor
\begin{equation}
  T(z)=\frac{1}{k-1}\Bigl(g_{ab}J^aJ^b+\varepsilon_{\alpha\beta}
  j^\alpha j^\beta \Bigr)(z) 
\end{equation}
gives the Virasoro algebra with the central charge $c=0$.

In particular from \eqref{ssl} one deduces the following
\begin{equation}
  T(z)=\beta\pd\gamma +\pd\xi\eta +\pd\txi\te -\oh\pd\varphi\pd\varphi
  -\oh\pd\phi\pd\phi(z)\quad.
\end{equation}

To complete the specification of the free field representation, let us
note that some free fields may be expressed in terms of the currents
\eqref{ssl} as

\begin{equation}\label{fssl}
  \beta(z)=J^-(z)\quad,\quad 
  \eta(z)=\sqrt 2j^--\oh\txi J^-(z)
  \quad,\quad
  \te(z)=\sqrt 2j^{\mm}-\oh\xi J^-(z)
  \quad.
\end{equation}


\section{The expectation value of $\gamma^n\gamma^{-n}$}
\renewcommand{\theequation}{B.\arabic{equation}}
\setcounter{equation}{0}

The purpose of this appendix is to compute the correlator
$\langle\gamma^n(z_1)\gamma^{-n}(z_2)\rangle_q$. There are a variety
of ways to do this. In general the technique of \cite{rp} allows us to
compute the correlator for any complex $n$ within the $(\beta,\gamma)$
system with $Q_{\text B}=0$. Of course, it can be generalized to
nonzero $Q_{\text B}$.  However, it seems more natural to make use of
bosonization in this case. If we begin with the bosonization recipe
\eqref{bos}, then $\gamma^n$ will be simply
\begin{equation}
  \gamma^n(z)=\ep^{n(i\sif -\sis )(z)}
  \quad.
\end{equation}
On the other hand, for the q-vacuum, the corresponding vertex operator
is given by
\begin{equation}
  V_q(z)=\ep^{ia\sif +b\sis (z)}\quad,\quad\text{with}\quad q=a+b
  \quad.
\end{equation}
We again use the Feigin-Fuchs representation where the background
charges are taken into account via screening operators. In general,
this technique is rather involved to be used in string theory because
it leads to integral representations of correlators but for the
problem at hand it works perfectly (one needs no screening operators).
The correlator is written as
\begin{equation}
  \langle\,\gamma^n(z_1)\gamma^{-n}(z_2)\,\rangle_q =
  \langle\,V_p^{\dagger}(\infty)
  \ep^{n(i\sif -\sis )(z_1)}
  \ep^{n(-i\sif +\sis )(z_2)}
  V_p(0)\,\rangle
  \quad,
\end{equation}
where $V_p^{\dagger}$ means the conjugate vacuum. A simple algebra
gives
\begin{equation}\label{gg}
  \langle\,\gamma^n(z_1)\gamma^{-n}(z_2)\,\rangle_q =
  \Big(\frac{z_2}{z_1}\Big)^{nq}
  \quad.
\end{equation}

To convince the reader, let us give an alternative derivation. It
follows from the definition of the q-vacuum that
$\langle\gamma^n(z_1)\gamma^{-n}(z_2) \rangle_q
=(\frac{z_2}{z_1})^{nq}\langle\gamma^n(z_1)\gamma^{-n}(z_2)\rangle_0$.
The last correlator is an analytic function of $z_1\,(z_2)$ in the
entire complex plane, so it is constant. To find this constant, note
that $\gamma^n(z_1)\gamma^{-n}(z_1)\equiv 1$. Thus we recover the
result \eqref{gg}.




\small
\begin{thebibliography}{99}
  
\bibitem{Mal} J. Maldacena, \ATMP{2}{1998} 231.
  
\bibitem{H} G. 't Hooft, ``Dimensional Reduction In Quantum Gravity'',
  in Salamfest 1993, p.284.  
  
\bibitem{LS} L. Susskind, J.Math.Phys. 36 (1995) 6377.
  
\bibitem{Green} M. B. Green, \NP{293}{1987} 593.
  
\bibitem{SUSY} E.  Witten and D. Olive, \PL{78}{1978} 97.
  
\bibitem{GKS} A. Giveon, D.  Kutasov and N. Seiberg, Comments on
  String Theory on $AdS_3$, hep-th/9806194.
  
\bibitem{Gaw} K. Gawedzki, \NP{328}{1989} 733.
  
\bibitem{I} K. Ito, Extended Superconformal Algebras on $AdS_3$,
  hep-th/9811002.
  
\bibitem{EFGT} S. Elitzur, O.Feinerman, A. Giveon and D. Tsabar,
  String Theory on $AdS_3\times S^3\times S^3\times S^1$,
  hep-th/9811245.
  
\bibitem{KLL} D. Kutasov, F. Larsen and R.  Leigh, String Theory in
  Magnetic Monopole Backgrounds, hep-th/9812027.
  
\bibitem{O} J. de Boer, H. Ooguri, H. Robins and J.  Tannenhauser,
  String Theory on $AdS_3$, hep-th/9812046.
  
\bibitem{Sug} K. Hosomichi and Y. Sugawara, Hilbert Space of
  Space-time SCFT in $AdS_3$ Superstring and $T^{4kp}/S_{kp}$ SCFT,
  hep-th/9812100.
  
\bibitem{YZ} M. Yu and B. Zhang, Light-Cone Gauge Quantization of
  String Theories on $AdS_3$ Space, hep-th/9812216.
  
\bibitem{FMS} D. Friedan,E. Martinec and S. Shenker, \NP{271}{1986}
  93.
  
\bibitem{W}
  M. Wakimoto, \CMP{104}{1986} 604;\\
  A.B. Zamolodchikov, Talk given at Montreal (1988), unpublished;\\
  M. Bershadsky and H. Ooguri, \PL{229}{1989} 374;\\
  P. Bowcock, R-L. K. Koktava and A. Taormina, \PL{388}{1996} 303; \\
  J. Rasmussen, \NP{510}{1998} 688.
  
\bibitem{rp} O. Andreev, \IJMP{10}{1995} 3221;\\
  J. L. Petersen, J. Rasmussen and M. Yu, \NP{457}{1995} 309.
  
\bibitem{V2} E. Verlinde and H. Verlinde, Lectures on String
  Perturbation Theory, Lectures at the Trieste Spring School on
  Superstrings, 1988.  
  
\bibitem{AF} O. Andreev and B. Feigin, \NP{433}{1995} 685.

\bibitem{cc}
  M. Li, \NP{382}{1992} 242; \\
  S. Kachru, \MPL{7}{1419}; \\
  Vl. S. Dotsenko, \MPL{7}{1992} 2505;\\
  J. L. F. Barb\' on, \IJMP{7}{1992} 7579.
  
\bibitem{VK} V. G.  Knizhnik, \CMP{112}{1987} 567.
  
\bibitem{DPL} L. J. Dixon, M. E.  Peskin and J. Lykken, \NP{325}{1989}
  329.
  
\bibitem{B} I. Bars, \PR{53}{1996} 3308.  
  
\bibitem{GKP} S. S. Gubser, I. R. Klebanov and A. M. Polyakov,
  \PL{428}{1998} 105.
  
\bibitem{M} E. Martinec, Matrix Models of AdS Gravity, hep-th/9804111.

\end{thebibliography}
\end{document}